\documentclass[showpacs,prl,twocolumn,amsmath,amssymb,superscriptaddress]{revtex4}
\usepackage{amsthm}
\usepackage{amssymb}
\usepackage{graphicx}
\usepackage{color}
\usepackage{cancel}
\usepackage{soul}
\definecolor{orange}{rgb}{1,0.5,0}

\begin{document}

\title{Genuine quantum and classical correlations in multipartite systems}

\author{Gian Luca Giorgi}
\affiliation{IFISC (UIB-CSIC),
Instituto de F\'isica Interdisciplinar y Sistemas Complejos, UIB Campus,
E-07122 Palma de Mallorca, Spain}
\author{Bruno Bellomo}
\affiliation{Dipartimento di Fisica, Universit\`a di Palermo, via Archirafi 36, IT-90123 Palermo, Italy}
\author{Fernando Galve}
\author{Roberta Zambrini}
\affiliation{IFISC (UIB-CSIC),
Instituto de F\'isica Interdisciplinar y Sistemas Complejos, UIB Campus,
E-07122 Palma de Mallorca, Spain}

\date{\today}
\begin{abstract}
Generalizing the quantifiers used to classify  correlations in bipartite systems, 
we define genuine total, quantum, and classical correlations in multipartite systems. 
The measure we give is based on the use of relative entropy to quantify the ``distance'' 
between two density matrices. Moreover, we show that, for pure states of three qubits, both quantum and classical bipartite correlations obey a 
ladder ordering law fixed by two-body mutual informations, or, equivalently, by one-qubit entropies.

\end{abstract}


\maketitle

Quantifying and characterizing the nature of correlations in a quantum state,
besides the fundamental scientific interest, has a crucial applicative importance
for the full development of quantum technologies~\cite{nielsen-chuang}. In this context,
the potential benefit present in multipartite instead of bipartite systems is a challenging and still largely open question~\cite{horod,ckw,osborne,kasz}.

The role of entanglement, first recognized as the characteristic trait of quantum mechanics~\cite{epr},
 has started to be debated, since there are evidences that
it does not capture all the quantum features of a system~\cite{info-no-ent}.
 In the case of bipartite systems,
the quantum discord $\cal{D}$, defined as
the difference between two quantum analogues of the classical mutual information~\cite{zurek,henderson}, has been widely accepted
as a fundamental tool due to its relevance in quantum computing tasks \textit{not} relying on entanglement~\cite{info-no-ent}.
Another approach to quantify the 
 correlations in an arbitrary $n$-partite system
is based on the distance between the system state  
and the closest states having the desired characteristics (e.g., separability, classicality) \cite{modi}.

In general, in the attempt of extending the concept of entanglement to the
multipartite case, definitions and results are less precise and there still exist many unsolved problems \cite{horod}.
Just to mention the most popular indicator, the role of  three-tangle as an entanglement measure is widely debated~\cite{ckw}.
An established feature of 
entanglement in multipartite systems is its monogamy, proved for qubits and continuous variables \cite{ckw,osborne,adesso}. 

In the case of quantum correlations, the problem of the extension to multipartite systems is poorly understood and there are many open issues.
A fundamental problem is the distinction between classical and quantum correlations.
Kaszlikowski  {\it et al.} introduced an axiomatic definition, based on the covariances of
local observables, to detect the presence of genuine $n$-partite classical correlations,
and found a state with genuine $n$-partite entanglement and no classical correlations according to such {\it ad hoc} indicator~\cite{kasz}.
It was then suggested that, in contrast with what happens in bipartite systems~\cite{mdms},
genuine quantum correlations can appear in states with no classical correlations.
The meaning of the adopted criterion has been discussed by Bennett {\it et al.},
who introduced a series of postulates that any {\it good} measure of multipartite
correlations should obey, and showed that the covariance is not suitable to be used to this end~\cite{grudka}. 

Recently, the problem of generalizing the definition of quantum discord in multipartite systems has been tackled following
different approaches~\cite{sarandy,chakrabarty}.

In this Letter, we approach this problem focusing on the introduction of a measure of {\it genuine} total, classical, and quantum correlations, based on
the use of the relative entropy to quantify the ``distance'' between states.
Genuine correlations will be defined as the amount of correlation that cannot be accounted for considering
any of the possible subsystems. For instance, merging two 
sub-parties without allowing them to cooperate,
should result in zero genuine correlations.

Let us recall the basic quantifiers for bipartite states.
Total correlations are measured by the mutual information ${\cal I}(\varrho_{a,b})= S(\varrho_a)+S(\varrho_b)-S(\varrho_{ab})$,
where $\varrho_j$ is the reduced density matrices of subsystem $j=a,b$ and $S(\varrho_j)=-{\rm Tr}\{\varrho_j \log \varrho_j\}$ is its von Neumann entropy.
According to Refs.~\cite{zurek,henderson}, classical
 correlations are given by ${\cal{J}}_{a:b}(\varrho)=\max_{\{E_i^b\}}\left[S(\varrho_a)-S(a|\{E_i^b\})\right]$,
where the  conditional entropy is  $S(a|\{E_j^b\})=\sum_ip_iS(\varrho_{a|E_i^b})$, $p_i={\rm
Tr}_{ab}(E_i^b\varrho)$, and where $\varrho_{a|E_i^b}= {\rm Tr}_b E_i^b\varrho/{p_i} $ is the density
matrix after a positive operator valued measure (POVM) $\{E_j^b\}$ has been performed on $b$.
Quantum discord is then defined as the difference between ${\cal I}$ and ${\cal{J}}$: $\delta_{a:b}(\varrho)=\min_{\{E_i^b\}}\left[S(\varrho_b)-S(\varrho_{ab})+S(a|\{E_i^b\})\right]$.
Both classical correlations and  discord are asymmetric under the exchange of $a$ and $b$, i.e.
${\cal J}_{a:b}\neq{\cal J}_{b:a}$ and ${\cal D}_{a:b}\neq{\cal D}_{b:a}$.
A possible symmetrization procedure, which we shall use in this Letter, consists of identifying classical
correlations as the maximum between the two values:  ${\cal J}(\varrho_{a,b})=\max[{\cal J}_{a:b},{\cal J}_{b:a}]$.
For the sake of consistency, quantum discord must be defined as ${\cal D}(\varrho_{a,b})=\min[{\cal D}_{a:b},{\cal D}_{b:a}]$.

Moving to the tripartite case, a possible extension of the Shannon classical mutual information is
${\cal I}_S=-\sum_{x,y,z}P_{x,y,z}\log_2 (P_{x}^{(a)} P_{y}^{(b)} P_{z}^{(c)})/P_{x,y,z}$,
where $x,y,z$ are the possible results of a measurement made on $a,b,c$, with respective probabilities $P_{x}^{(a)}, P_{y}^{(b)}, P_{z}^{(c)}$, being
$ P_{x,y,z}$ the probability of a collective measurement.
In the quantum case, by replacing the classical probability distributions
 by the appropriate density matrices and the Shannon entropy  by the von Neumann entropy,
the total information (or correlation information) of a tripartite state $\varrho_{abc}\equiv\varrho$ is given by
\begin{equation}
 T(\varrho)=S(\varrho_a)+S(\varrho_b)+ S(\varrho_c)-S(\varrho),\label{total}
\end{equation}
As shown in Ref.~\cite{modi},
$T(\varrho)$ is the distance, as measured by the relative entropy, between $\varrho$ and its closest product state 
that does not contain any correlations $\pi_\varrho=\varrho_a\otimes\varrho_b\otimes\varrho_c$:
$T(\varrho)=S(\varrho\parallel\pi_\varrho)={\rm Tr}[ \varrho (\log\varrho -\log\pi_\varrho)]$.

Having defined  total correlations, we are left to find how much of them is quantum and how much is classical.
Using Bayes' rules, which allow to write $P_{x,y,z}=P_{x|y,z}P_{y|z}P_{z}$ and the other permutations, 
we find out a series of $6$ equivalent versions of the total Shannon classical information ${\cal I}_S$. 
When extended to the quantum formalism, for each choice of the indices  one must look for the complete measurement on 
two sub-parties maximizing the quantum version of ${\cal I}_{S}$. 
This leads to ${\cal J}_{i:j:k}(\varrho)=S(\varrho_j)-S(\varrho_{j|i})+S(\varrho_k)-S(\varrho_{k|ji})$, 
where $S(\varrho_{j|i})=\min_{\{E_l^i\}}\left[S(j|\{E_l^i\})\right]$ and 
$S(\varrho_{k|ji})=\min_{\{E_l^i,E_m^j\}}\left[S(k|\{E_l^i,E_m^j\})\right]$ being $\{E_l^i\}$ and $\{E_m^j\}$ POVM's  on parties $i$ and $j$.
Using the same symmetrization principle employed for bipartite states, we define total classical correlations as the maximum among the $6$ indices permutations $p_{\{i,j,k\}}$:
\begin{equation}
 {\cal J}(\varrho)=\max_{p_{\{i,j,k\}}} [S(\varrho_j)-S(\varrho_{j|i})+S(\varrho_k)-S(\varrho_{k|ji})].\label{totalJ}
\end{equation}

According to Ref.~\cite{modi}, ${\cal J}(\varrho)$ measures the distance between the  classical state  closest to $\varrho$ and its closets product state.
${\cal J}(\varrho)$  is
 the sum of two independent classical correlations: ${\cal J}(\varrho_{k|ji})=S(\varrho_k)-S(\varrho_{k|ji})$ is the bipartite classical correlation
between $k$ and $ji$, and
 ${\cal J}(\varrho_{j|i})=S(\varrho_j)-S(\varrho_{j|i})$ is the classical correlation
between $j$ and $i$. Both ${\cal J}(\varrho_{k|ji})$ and ${\cal J}(\varrho_{j|i})$ are
relative entropies~\cite{modi}.
Also the discord could be related to  relative entropies~\cite{modi}, but 
in order to simplify the presentation, we call total quantum discord ${\cal D}(\varrho)=T(\varrho)-{\cal J}(\varrho)$, in agreement with the original definition~\cite{zurek,henderson}.

Our goal is to find a decomposition for $T,{\cal J}$, and ${\cal D}$  such that they can be written as the sum of
genuinely tripartite correlations and a contribution deriving from any possible partitions. 
Genuine correlations should contain all the contributions that cannot be accounted for considering
any of the possible subsystems. 
As stated in Ref.~\cite{grudka}, a state of $n$ particles has genuine $n$-partite correlations if it is nonproduct in every bipartite cut.

Building on this criterion, here we define
genuine tripartite correlations $T^{(3)}(\varrho)$ as the difference between $T(\varrho)$ and the maximum among the bipartite correlations:
\begin{equation}
  T^{(3)}(\varrho)=T(\varrho)-T^{(2)}(\varrho),\label{trip}
\end{equation}
where $T^{(2)}(\varrho)=\max[{\cal I}(\varrho_{a,b}),{\cal I}(\varrho_{a,c}),{\cal I}(\varrho_{b,c})]$. If we assume
${\cal I}(\varrho_{a,b})\ge {\cal I}(\varrho_{a,c})\ge {\cal I}(\varrho_{b,c})$, we have $ T^{(3)}(\varrho)=S(\varrho_{ab})+S(\varrho_{c})-S(\varrho)={\cal I}(\varrho_{ab,c})$.
Then, total genuine correlations coincide with the lowest bipartite mutual information present in the state.

$ T^{(3)}(\varrho)$ has a direct interpretation in the approach according
to which a quantifier for a given property is equal to the distance
between $\varrho$ and the closest    state without that property.
In fact, 
in terms of relative entropy, $ T^{(3)}$ measures the distance between $\varrho$ and the closest state with no tripartite correlations:

\indent {\it Theorem 1.  {\bf ---}} Given the definition (\ref{trip}), $T^{(3)}(\varrho)=\min[S(\varrho\parallel \varrho_{ab}\otimes\varrho_{c}),S(\varrho\parallel \varrho_{ac}\otimes\varrho_{b}),S(\varrho\parallel \varrho_{bc}\otimes\varrho_{a})]$.

\noindent Here $\varrho_{ij}$ and $\varrho_{k}$ are, respectively, the two-party and one-party reduced density matrices of $\varrho$. 
The proof is postponed to the end of the Letter.
Of course, also $T^{(2)}(\varrho)$ can be thought in terms of distance between states: $T^{(2)}(\varrho)=S(\varrho_{ij}||\varrho_i \otimes \varrho_j)$.

Our quantifier of Eq.~(\ref{trip})
is compatible with the definition of genuine correlations given  in Ref.~\cite{grudka}. We show this by demonstrating that if tripartite
correlations are zero, then the total state is product at least along a bipartite cut.
Indeed, if  ${\cal I}(\varrho_{ab,c})=0$, the total state is factorized at least along this bipartite cut, because
${\cal I}(\varrho_{ab,c})=0$ implies $\varrho=\varrho_{ab}\otimes\varrho_{c}$.

Next step is the quantification of genuine classical (${\cal J}^{(3)}$) and quantum (${\cal D}^{(3)}$) correlations. 
First, we observe that they can be evaluated considering that $T^{(3)}(\varrho)$ is actually a bipartite mutual information
and can be divided into its classical and quantum parts, as for ordinary bipartite states~\cite{zurek,henderson}. 
This would then lead to a first definition of ${\cal D}^{(3)}$ and ${\cal J}^{(3)}$.
For instance, for pure states, we have ${\cal D}^{(3)}={\cal J}^{(3)}=T^{(3)}/2=\min_i[S(\varrho_{i})]$.

On the other hand, to pursue our goal, and consistently with the definition of $T^{(3)}$,
we identify genuine classical and quantum  correlations as the
difference between their total counterparts and the maximum among bipartite correlations:
\begin{eqnarray}
 {\cal J}^{(3)}(\varrho)&=&{\cal J}(\varrho)-{\cal J}^{(2)}(\varrho),\nonumber\\
 {\cal D}^{(3)}(\varrho)&=&{\cal D}(\varrho)-{\cal D}^{(2)}(\varrho),\label{ridef}
\end{eqnarray}
where $ {\cal J}^{(2)}(\varrho)=\max[{\cal J}(\varrho_{a,b}),{\cal J}(\varrho_{a,c}),{\cal J}(\varrho_{b,c})]$ and
 ${\cal D}^{(2)}(\varrho)=\min[{\cal D}(\varrho_{a,b}),{\cal D}(\varrho_{a,c}),{\cal D}(\varrho_{b,c})]$.

Since, in the last two paragraphs, we have introduced  $ {\cal J}^{(3)}$ and $ {\cal D}^{(3)}$ in two different ways, 
we must show that the two definitions coincide. 
So far, we discussed the general problem of a tripartite state without specifying the nature of the system. 
From now on, we specialize on the case of pure states of three qubit, where we are able to get analytical results. 
One can prove that, if  $ \varrho$ is pure, the measurement performed on ${\{ij\}}$ that minimizes $S(\varrho_{k|ji})$ 
is made by local measurements on $i$ and on $j$ [this gives $S(\varrho_{k|ji})=0$ ] and that the local measurement on $i$ is the same minimizing $S(\varrho_{j|i})$.
Henceforward, we shall make the following assumption:
\begin{equation}
 {\cal I}(\varrho_{a,b})\ge {\cal I}(\varrho_{a,c})\ge {\cal I}(\varrho_{b,c}).\label{ass}
\end{equation}
We need to prove the following statements:

{\it Lemma. {\bf ---}} Given a pure state of three qubits, under  assumption (\ref{ass}),
\begin{equation}
 S(\varrho_a)+{\cal E}(\varrho_{b,c}) \le S(\varrho_b) + {\cal E}(\varrho_{a,c})\le S(\varrho_c)+{\cal E}(\varrho_{a,b}) ,\label{conve}
\end{equation}
where ${\cal E}$ is the entanglement of formation.

{\it Theorem 2. {\bf ---}} Given a pure state of three qubits, with assumption (\ref{ass}), 
the equivalent chain of inequalities is found for classical correlations:
$ {\cal J}(\varrho_{a,b})\ge{\cal J}(\varrho_{a,c})\ge{\cal J}(\varrho_{b,c})$. Moreover,
 $ {\cal D}(\varrho_{a,b})\ge\max[{\cal D}(\varrho_{a,c}),{\cal D}(\varrho_{b,c})]$.

The proofs are given at the end of the Letter. This theorem allows one to calculate all the quantifiers we have defined. 
Noticing that, 
since the total $\varrho$  is pure, all the relative entropies $S(\varrho_{i|jk})$ are zero,
we find the following total classical and quantum correlations:
\begin{eqnarray}\label{cor1}
 {\cal J}(\varrho)&=&S(\varrho_b)+S(\varrho_c)-{\cal E}(\varrho_{b,c}),\nonumber\\
{\cal D}(\varrho)&=&S(\varrho_a)+{\cal E}(\varrho_{b,c}),
\end{eqnarray}
while, the corresponding bipartite correlations are
\begin{eqnarray}\label{cor2}
 {\cal J}^{(2)}(\varrho)&=&S(\varrho_b)-{\cal E}(\varrho_{b,c}),\nonumber\\
{\cal D}^{(2)}(\varrho)&=&S(\varrho_a)-S(\varrho_c)+{\cal E}(\varrho_{b,c}).
\end{eqnarray}
As for genuine correlations, we have, in agreement with the first definition,
\begin{equation}\label{cor3}
 {\cal J}^{(3)}(\varrho)={\cal D}^{(3)}(\varrho)=S(\varrho_c).
\end{equation}
Then, genuine classical and quantum correlations are equal.
This result is reminiscent of the well known equality between ${\cal D}$ and ${\cal J}$ for bipartite pure states.

Recalling the procedure employed to obtain ${\cal J}(\varrho)$ through the application of Bayes' rules, we identify ${\cal J}(\varrho_{c|ba})$ with $ {\cal J}^{(3)}(\varrho)$ and
${\cal J}(\varrho_{b|a})$ with $ {\cal J}^{(2)}(\varrho)$.

{\it Corollary. {\bf ---}}
${\cal J}^{(3)}(\varrho)$ is the minimum distance between the classical state closest to $\varrho$ 
and a state which is product along a bipartite cut.

This corollary, whose proof is postponed, concludes our generalization of correlations to tripartite systems.

Let us compare our indicator of genuine quantumness ${\cal D}^{(3)}$ with the three-tangle 
$\tau_{abc}$~\cite{ckw}. For a generic pure state $|\psi\rangle=\lambda_0 |0,0,0\rangle+\lambda_1 e^{i \theta} |1,0,0\rangle+\lambda_2 |1,0,1\rangle+\lambda_3 |1,1,0\rangle+\lambda_4 |1,1,1\rangle$ \cite{acin}, $\tau_{abc}=4\lambda_0\lambda_4$. The subfamily of W states, obtained fixing $\lambda_4=0$ 
has then zero three-tangle. 
This last point invalidates the use $\tau_{abc}$ as a measure of genuine quantum correlations. 
 On the other hand, genuine quantum correlations present in W states are captured by ${\cal D}^{(3)}$, which can be used as a proper quantifier of quantumness of all possible states.
Among this family, the maximum is ${\cal D}^{(3)}\simeq 0.918$ and is obviously reached by $|\psi_{W}\rangle=(|0,0,1\rangle+|0,1,0\rangle+|1,0,0\rangle)/\sqrt{3}$.
Furthermore, the maximum value for a generic state, ${\cal D}^{(3)}=1$, is achieved, as expected, by the maximally entangled state $|\psi_{GHZ}\rangle=(|0,0,0\rangle+|1,1,1\rangle)/\sqrt{2}$. 

Next, we quantify the description of a transition to a completely classical product state for these extremal states. Namely, we consider the states
 $|\tilde\psi_{GHZ}(p)\rangle=\sqrt{p}|\psi_{GHZ}\rangle+\sqrt{1-p}|1,0,0\rangle$ and
$|\tilde\psi_{W}(p)\rangle=\sqrt{p}|\psi_{W}\rangle+\sqrt{1-p}|0,0,0\rangle$, with $p\in[0,1]$, and calculate $T,{\cal D},{\cal J},$ and ${\cal D}^{(3)}(={\cal J}^{(3)})$ as a function of $p$.
 While $|\tilde\psi_{GHZ}(p)\rangle$ has always more overall correlations and more genuine correlations, for $p\gtrsim 0.75$, $|\tilde\psi_{W}(p)\rangle$ is more quantum, being the correlation in $|\tilde\psi_{GHZ}(p)\rangle$ dominated by its classical part. 
Furthermore, in the case of $|\tilde\psi_{W}(p)\rangle$, ${\cal D}$ is always greater than ${\cal J}$, while the opposite is true for $|\tilde\psi_{GHZ}(p)\rangle$.

In analogy with the case of entanglement, it is interesting to discuss the problem 
of the existence of equivalence classes for our quantifiers. We observe that all
the quantities appearing in Eqs.~(\ref{cor1},\ref{cor2},\ref{cor3}) are invariant 
under local unitary operations. For example, the pure state of three
qubits with maximum amount of genuine quantum correlations would be
$U_i|\psi_{GHZ}\rangle\langle\psi_{GHZ}|U_i^\dagger$ where $U_i$ is a
local unitary acting on any of the one-qubit subsystems, thus preserving
${\cal D}^{(3)}=\min_i[S(\varrho_i)]=1$. A more general prescription
would be to find the Bloch representation of a given state and define
its equivalence class as all (pure) states sharing the same one-particle
Bloch vector having the maximum length. 
Non-unitary channels would map $\varrho$ into a mixed state, 
and Eqs.~(\ref{cor1},\ref{cor2},\ref{cor3}) would not hold anymore.
Nevertheless, genuine total correlations here defined are  the smaller
bipartite correlations between two sub-parties and the third one.
Then, the action of local noisy channels could result in a increase of ${\cal D}^{(3)}$ 
as it may happen to bipartite quantum correlations~\cite{noisediscord}.

Part of the results of this Letter can be extended to general $n$-partite systems. Total correlations
are  $T(\varrho)=\sum_{i=1}^n S(\varrho_n)-S(\varrho)$, and their genuine part is measured by the relative entropy between $\varrho$
and the closest state without $n$-partite correlations, i.e. the closest state which is factorized at least along a bipartite cut.
Then, $T^{(n)}$ is still the smaller bipartite mutual information, and the considerations made before apply.
In particular, it is still possible to write down, for pure states, ${\cal D}^{(n)}={\cal J}^{(n)}=T^{(n)}/2$ and calculate them as the minimum  entropy
over all the possible reduced $k$-partite ($k<n$) states. 
Alternatively, we can define,  in a more complicated way, through a ladder procedure, ${\cal D}^{(n)}$ and ${\cal J}^{(n)}$ in analogy with Eq.~\ref{ridef}.
On the other hand, the invariance of the tangle under index permutations is known only for three-qubit states.
Therefore, beyond the tripartite case, it is not evident how to generalize Theorem 2, even if its validity seems reasonable.

In conclusion, we have shown that, given a multipartite pure state of three qubits, classical correlations and quantum discord of all
the possible bipartitions obey a hierarchical law fixed by one-qubit entropies. This allows us to formulate
a definition of genuine correlations based on the use of relative entropies.

\indent {\it Proof of Theorem 1. {\bf ---}}
 Let us recall that 
$ T^{(3)}(\varrho)=\min_i[{\cal I}(\varrho_{jk,i})]$.
We observe that ${\cal I}(\varrho_{ab,c})$ measures the distance between $\varrho$ 
and the closest state 
of the form $\tilde\varrho_{ab}\otimes\tilde\varrho_{c}$,
 where $\tilde{\varrho}_{ab}$ and $\tilde{\varrho}_{c}$ are respectively arbitrary 
two-party and one-party states. 
As proved in Ref.~\cite{modi}, the minimum distance occurs when $\tilde{\varrho}_{ab}$ 
and $\tilde{\varrho}_{c}$ are 
the marginals of the total state $\varrho$, that is
${\cal I}(\varrho_{ab,c})=S(\varrho\parallel\varrho_{ab}\otimes\varrho_{c})$. 
Taking the minimum among the three possible 
cuts is then equivalent to finding the distance from the closest state with no tripartite correlations. 
$\square$

\indent {\it Proof of Lemma. {\bf ---}}
 First of all, since the total state is pure, assumption (\ref{ass}) is equivalent to
$S(\varrho_i) \ge S(\varrho_j)\ge S(\varrho_k)$. Furthermore, 
as shown in Ref.~\cite{winter}, for pure tripartite qubit states, the following relationships
between conditional entropies and entanglement of formation (${\cal E}$) applies: $S(\varrho_{i|j})=S(\varrho_{k|j})={\cal E}(\varrho_{i,k})$.
The use of this formula to calculate quantum discord has been introduced  by Fanchini {\it et al.}~\cite{fanchini}. 
Then, we have ${\cal J}_{i:j}(\varrho)=S(\varrho_i)-{\cal E}(\varrho_{i,k})$.

Upper and lower bounds for the entanglement of formation can be obtained using the results by
Coffman, Kundu, and Wootters~\cite{ckw}:
\begin{equation}
{\cal C}^2_{a}+{\cal C}^2_{b,c}={\cal C}^2_{b}+{\cal C}^2_{a,c}={\cal C}^2_{c}+{\cal C}^2_{a,b}.\label{c3}
\end{equation}
Here ${\cal C}_{i,j}$ is the concurrence, the entanglement monotone 
defined in Ref.~\cite{wootters}, between $i$ and $j$, and ${\cal C}_{i} =2 \sqrt{\det \varrho_i}$ is the concurrence between $i$ and $jk$.
Thus, if $S(\varrho_a) \ge S(\varrho_b)\ge S(\varrho_c)$, then ${\cal C}_{a} \ge {\cal C}_{b}\ge {\cal C}_{c}$.
The entanglement distribution among many parties 
implies ${\cal C}^2_{i}\ge{\cal C}^2_{i,j}+{\cal C}^2_{i,k}$~\cite{ckw}.
Concurrence and  entanglement of formation are related by ${\cal E}=h[(1+\sqrt{1-{\cal C}^2})/2]$,
where $h$ is the binary entropy $h(x)=-x\log_2 x-(1-x)\log_2(1-x)$, and
${\cal E}(\varrho_i)=S(\varrho_i)$.
Both ${\cal E}$ and ${\cal C}$ admit values between $0$ and $1$,
and ${\cal E}$ is a concave function of ${\cal C}^2$. 
Then applying, the function ${\cal E}$ to all elements of Eq. (\ref{c3}), and noticing that
$h((1+\sqrt{1-x})/2)+h((1+\sqrt{1-L+x})/2)$ ($0\leq L\leq 1$ represents ${\cal C}^2_{i}+{\cal C}^2_{j,k}$)
has a maximum for $x=L/2$ and two minima for $x=0$ and $x=L$, we obtain~(\ref{conve}).
$\square$

\indent {\it Proof of Theorem 2. {\bf ---}}
Inequalities ~(\ref{conve})
 imply ${\cal J}_{a:b}\le {\cal J}_{b:a}$, ${\cal J}_{a:c}\le {\cal J}_{c:a}$, and
${\cal J}_{b:c}\le {\cal J}_{c:b}$, or, in other words, ${\cal J}(\varrho_{a,b})={\cal J}_{b:a}$, ${\cal J}(\varrho_{a,c})={\cal J}_{c:a}$,
and  ${\cal J}(\varrho_{b,c})={\cal J}_{c:b}$. Furthermore, it is also  immediate to verify that
${\cal J}(\varrho_{a,b})\ge{\cal J}(\varrho_{a,c})\ge{\cal J}(\varrho_{b,c})$.

As for the second thesis of the theorem, using the purity of the state and the results of Ref.~\cite{winter}, the bipartite discord is 
${\cal D}_{i:j}(\varrho)=S(\varrho_j)-S(\varrho_k)+{\cal E}(\varrho_{i,k})$.
With the help of inequalities~(\ref{conve}), we can immediately state that
${\cal D}(\varrho_{a,b})={\cal D}_{b:a},{\cal D}(\varrho_{a,c})={\cal D}_{c:a}$ and ${\cal D}(\varrho_{b,c})={\cal D}_{c:b}$.
It is also evident that ${\cal D}(\varrho_{a,b})\ge {\cal D}(\varrho_{a,c})$, since
${\cal D}_{b:a}-{\cal D}_{c:a}=S(\varrho_b)-S(\varrho_c)\ge 0$ by assumption.

To end our proof, we show that ${\cal D}_{b:a}\ge{\cal D}_{c:b}$, or
\begin{equation}
2S(\varrho_a)-S(\varrho_b)-S(\varrho_c)\ge [{\cal E}(\varrho_{a,c})-{\cal E}(\varrho_{b,c})].\label{toproof}
\end{equation}
 First, 
 due to Eq.~(\ref{c3}), ${\cal E}(\varrho_{a,c})-{\cal E}(\varrho_{b,c})$ decreases
by decreasing 
$S(\varrho_a)-S(\varrho_b)$. Furthermore, due to the
concavity of ${\cal E}({\cal C}^2)$
the smaller $S(\varrho_a)-S(\varrho_b)$ is, the smaller is the derivative of ${\cal E}(\varrho_{a,c})-{\cal E}(\varrho_{b,c})$.
This implies that such a function is monotone under the change of $S(\varrho_b)$ for a fixed value of $S(\varrho_a)$. 
Then, we vary $S(\varrho_b)$ from its minimum value allowed [$S(\varrho_a)/2$], where inequality~(\ref{toproof}) is not violated, 
up to the maximum $S(\varrho_b)$ where~(\ref{toproof}) becomes an equality. 
This is true independently on the value of $S(\varrho_c)$, which can be taken as a static parameter. Then, due to the monotonicy of 
${\cal E}(\varrho_{a,c})-{\cal E}(\varrho_{b,c})$, we conclude that~(\ref{toproof}) cannot be violated. $\square$

\indent {\it Proof of Corollary. {\bf ---}}
According to Ref.~\cite{modi}, ${\cal J}^{(3)}(\varrho)$ is nothing else than the minimum ``distance'' (as measured by the relative entropy)
between $\chi_{\varrho}$ (the classical state closest to $\varrho$) and states that can be written as  $\varrho_{ab}\otimes\varrho_c$, as mentioned above. To see it, it is sufficient to observe that
if we have measured such a distance along different bipartite cuts, we would find as a result $S(\varrho_a)$ or $S(\varrho_b)$,
which are assumed to be greater than $S(\varrho_c)$. $\square$

\acknowledgments{Funding from FISICOS (FIS2007-60327), CoQuSys (200450E566), 
"Acci\'on Especial" CAIB (AAEE0113/09) projects and from  `Juan de la Cierva'' and ``Junta para la Ampliaci\'on de Estudios'' programs are  acknowledged. 
}


\begin{thebibliography}{10}

\bibitem{nielsen-chuang} M. A. Nielsen and I. L. Chuang, {\it Quantum Computation and Quantum Information}, Cambridge University Press, Cambridge (2000).

\bibitem{horod} R. Horodecki  {\it et al.},
Rev. Mod. Phys. {\bf 81}, 865 (2009).

\bibitem{ckw} V. Coffman  {\it et al.}, Phys. Rev. A {\bf 61}, 052306 (2000).

\bibitem{osborne} T. J. Osborne and F. Verstraete, Phys. Rev. Lett. {\bf 96}, 220503 (2006).




\bibitem{kasz} D. Kaszlikowski  {\it et al.}, Phys. Rev. Lett. {\bf 101}, 070502 (2008).

\bibitem{epr} A. Einstein {\it et al.}, Phys. Rev. {\bf 47}, 777 (1935).


\bibitem{info-no-ent} E. Knill and R. Laflamme, Phys. Rev. Lett. {\bf 81}, 5672 (1998);
A. Datta {\it et al.}, {\it ibid.}  {\bf 100}, 050502 (2008);
B. P. Lanyon  {\it et al.}, {\it ibid.}   {\bf 101}, 200501 (2008);


\bibitem{zurek}  H.  Ollivier and W.  H.  Zurek,  Phys. Rev. Lett.  {\bf 88}, 017901 (2001).


\bibitem{henderson} L.  Henderson and V.  Vedral,  J. Phys. A  {\bf 34}, 6899 (2001).

\bibitem{modi} K. Modi  {\it et al.}, Phys. Rev. Lett. {\bf 104}, 080501 (2010).


\bibitem{adesso} T. Hiroshima  {\it et al.}, Phys. Rev. Lett. {\bf 98}, 050503 (2007). 



\bibitem{mdms}  A. Ferraro  {\it et al.}, Phys. Rev. A {\bf 81} (2010)  052318; F. Galve  {\it et al.}, {\it ibid.} {\bf 83}, 012102 (2011).

\bibitem{grudka} C. H. Bennett {\it et al.}, Phys. Rev. A {\bf 83}, 012312 (2011).

\bibitem{sarandy}  C. C. Rulli and M. S. Sarandy, arXiv:1105.2548.

\bibitem{chakrabarty}  I. Chakrabarty {\it et al.},  arXiv:1006.5784.



\bibitem{winter} M. Koashi and A. Winter,  Phys. Rev. A  {\bf 69}, 022309 (2004).

\bibitem{fanchini} F. F. Fanchini {\it et al.}, Phys. Rev. A {\bf 84}, 012313 (2011).

\bibitem{wootters}  W. K. Wootters, Phys. Rev. Lett. {\bf 80}, 2245 (1998).

\bibitem{acin} A. Ac\'in {\it et al.}, Phys. Rev. Lett. {\bf 85}, 1560 (2000);
A. Ac\'in {\it et al.},  {\it ibid.}  {\bf 87}, 040401 (2001).

\bibitem{noisediscord} S. Campbell {\it et al.}, arXiv:1105.5548;
F. Ciccarello and V. Giovannetti, arXiv:1105.5551; 
A. Streltsov {\it et al.}, arXiv:1106.2028.

\end{thebibliography}
\end{document}